\documentclass{mscs}
\usepackage{graphicx}
\usepackage{bm}
\usepackage{amsmath}
\usepackage{amsfonts}
\usepackage{amssymb}
\usepackage[utf8]{inputenc}
\usepackage[american]{babel}

\newcommand{\mb}[1]{\mathbf{#1}}

\newcommand{\ket}[1]{|#1\rangle}
\newcommand{\bra}[1]{\langle#1|}

\newcommand{\scalar}[2]{\langle#1|#2\rangle}
\newcommand{\op}[1]{|#1\rangle\langle#1|}

\begin{document}

  \title[Spatial search on a honeycomb network]{Spatial search on a honeycomb network}
  \author[G. Abal, R. Donangelo, F. Marquezinho and R. Portugal] {G. Abal$^1$\thanks{corresponding author: abal@fing.edu.uy}, R. Donangelo$^1$, F.L.~Marquezino$^2$ and R.~Portugal$^2$\\
  $^1$Instituto de F\'{\i}sica, Facultad de Ingenier\'{\i}a, UdelaR, \addressbreak
  C.C. 30, C.P. 11300, Montevideo, Uruguay\\
  $^2$Laborat\'{o}rio Nacional de Computa\c{c}\~{a}o Cient\'{\i}fica - LNCC, \addressbreak
  Av.  Get\'{u}lio Vargas 333, Petr\'{o}polis, RJ, 25651-075, Brazil}

\begin{abstract}
The spatial search problem consists in minimizing the number of steps required to find a given site in a
network, under the restriction that only oracle queries or translations to neighboring sites are allowed. We propose a quantum algorithm for the spatial search problem on a honeycomb lattice with $N$ sites and
torus-like boundary conditions. The search algorithm is based on a modified quantum walk on an hexagonal lattice and the general framework proposed by Ambainis, Kempe and Rivosh~\cite{AKR05} is employed to show that the time complexity of this quantum search algorithm is $O(\sqrt{N \log N})$.
\end{abstract}

\maketitle

\section{Introduction}

Quantum Walks (QW) are useful tools to generate new quantum algorithms ~\cite{Amb03,SKW03,AKR05}.
For example, the optimal algorithm for solving the element distinctness problem, which aims to determine whether a set has repeated elements or not, is based on QWs~\cite{Amb03}. An optimal search algorithm   equivalent to the celebrated Grover's algorithm ~\cite{Gro96a}, uses a modified QW on an $n$-dimensional hypercube to find an element among $N$ sites after $O(\sqrt{N})$ steps \cite{SKW03}. Although the QW is a unitary (i.e. invertible) process, it is often introduced as the quantum analog of a random walk or, more generally, of a Markov process. There are two versions of QWs: discrete-time~\cite{ADZ93} and continuous-time~\cite{FG98} walks. The first one uses an auxiliary Hilbert space, which plays the role of a quantum ``coin'' whose states determine the directions of motion.  Even though both types of QW's have similar dynamics, they are not equivalent. For instance, the optimal algorithm for spatial search in two-dimensional grids using the continuous-time version has no advantage over the classical algorithm in terms of time complexity~\cite{CG04}, while the algorithm based on the discrete-time version has an almost quadratic improvement~\cite{Tulsi08}.

Grover's algorithm applies to non-ordered databases, where there is no notion of distance between two  elements.
However, when storing information in physical memory, a given item is stored at a specific location.
This poses an interesting alternative version of searching, called spatial search, as the problem of finding a
marked location in a rigid structure using only local operations: in one time step one can either query an oracle
for the given site or move to a neighbouring site. Benioff~\cite{Ben02} addressed this problem on a two-dimensional
square lattice with $N$ points. He was the first to point out that a straightforward application of Grover's algorithm
with the spatial search constrain requires $\Omega(N)$ steps with no improvement over classical
algorithms in terms of time complexity. Aaronson and Ambainis~\cite{AA03} have developed a quantum algorithm for
this problem with time complexity $O(\sqrt N \log^2 N)$. Ambainis, Kempe and Rivosh (AKR)~\cite{AKR05} have proposed
a QW-based algorithm which improves the time complexity to $O(\sqrt N \log N)$. Recently, Tulsi~\cite{Tulsi08} has
proposed an improved version of the AKR spatial search algorithm for two-dimensional square lattices, with a time
complexity of $O(\sqrt{N\log N})$. It is an open problem whether the lower bound $\Omega(\sqrt N)$ can be achieved
for the spatial search on two-dimensional lattices~\cite{BBBV}. AKR have proposed a generalized framework for QW-based algorithms on lattices of arbitrary structure, in which the time-complexity of the algorithm may be obtained from the eigenvalue spectrum of the QW evolution operator. Following AKR, we shall refer to this formalism as the \textit{abstract search framework}.

In this paper, we provide a new QW-based algorithm which solves the spatial search problem in a hexagonal (honeycomb)
network in $O(\sqrt{N\log N})$ steps. The time complexity is analyzed using the abstract search framework just discussed.
The hexagonal network has received attention from condensed matter physicists for many years, due to its role in the
band theory of graphite~\cite{Wallace}. More recently, the development of graphenes (two-dimensional hexagonal arrays
of Carbon atoms) and its possible uses in quantum computation ~\cite{NMDB06} have renewed the interest on these networks~\cite{GM07}. The paper is organized as follows. In Section~\ref{sec:hexnet} we discuss the implementation of a quantum walk on a periodic hexagonal network and obtain the evolution operator in the Fourier-transformed space. In Section~\ref{sec:order}
we summarize the abstract search framework and use it to evaluate the time complexity of the search algorithm on a 
hexagonal lattice. In Section~\ref{sec:conc} we present our conclusions.

\section{QW on the hexagonal network}
\label{sec:hexnet}

The Hilbert space of a QW, ${\cal H}={\cal H}_C\otimes {\cal H}_P$ is composed of a coin, ${\cal H}_C$, and a position
subspace, ${\cal H}_P$. The evolution operator is of the form $U=S \cdot (C\otimes I)$ where $C$ is a unitary operation
in ${\cal H}_C$, $I$ is the identity in ${\cal H}_P$ and $S$, a shift operation in ${\cal H}$, performs a conditional
one-step displacement as determined by the current coin state. The main challenge to obtain the time complexity of a
QW-based algorithm on a honeycomb lattice is the calculation of the spectral decomposition of the evolution operator $U$
of the underlying QW. The abstract search framework is based on a modified evolution operator $U^\prime=S \cdot C^\prime$,
obtained from the standard quantum walk operator $U$ by replacing the coin operation $C$ with a new unitary operation $C^\prime$ which is not restricted to ${\cal H}_C$ and acts differently on the searched vertex. Ambainis and coworkers have shown that the time complexity of the spatial search algorithm can be obtained from the spectral decomposition of the evolution operator $U$ of the unmodified QW ~\cite{AKR05}, which is usually simpler than that of $U'$.

\begin{figure}
  \centering
  \includegraphics[scale=0.7]{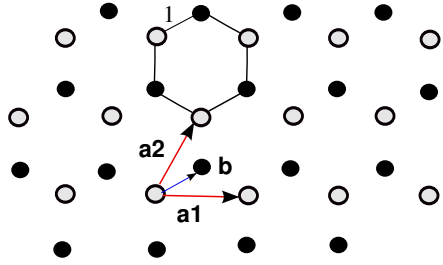}
  \caption{Elementary vectors for the honeycomb network. The white sites form a lattice and black sites form the associated basis.}
  \label{fig:lattice}
\end{figure}

In regular networks, the use of the Fourier transform on the spatial coordinates considerably simplifies the expressions for the eigenvalues and eigenvectors. It is known that a Bravais lattice has an associated reciprocal lattice~\cite{Kittel} and this provides a systematic way for obtaining the Fourier transform. The honeycomb network is not a Bravais lattice, but this
can be circumvented by splitting the vertices into two sets with $N/2$ sites each (the lattice and basis sets) and encoding
the which-set information on an auxiliary one-qubit state. In Fig.~\ref{fig:lattice}, we distinguish between the $N/2$ lattice
sites (gray) and the $N/2$ basis sites (black) using a color code.

Let us consider the distance between two adjacent sites of
the hexagonal network as the unit distance. Then, the vectors $\mb{a_1}$ and $\mb{a_2}$ which connect two neighboring lattice
sites (see Fig.~\ref{fig:lattice}) have norm $\sqrt{3}$ and span an angle of $60^o$. The unit vector $\mb{b}$ which locates
the basis site adjacent to a given lattice site is given by $\mb{b}=\frac13 (\mb{a_1}+\mb{a_2})$. An arbitrary lattice point
may be addressed by a vector with integer components
\begin{equation}\label{eq:lattice}
  \mb{r}=n_1\mb{a_1}+n_2\mb{a_2}
\end{equation}
and each lattice point has an associated basis point at
$\mb{r}+\mb{b}$.
Assume periodicity in both directions (von Karmann boundary conditions), so that $n_1,n_2\in\left[0, m-1\right]$. For simplicity, we  consider a number of sites $N$ such that $N=2m^2$, for some integer $m$.  Thus, for an $N$-element network, we have $N/2$
kets $\ket{n_1,n_2}$ spanning the position subspace associated to the lattice. The $N/2$ basis sites are accounted for by introducing an auxiliary qubit, $\{\ket{0},\ket{1}\}$, which is zero for a lattice site and 1 for a basis site. Thus $\ket{0; n_1,n_2}\equiv\ket{0}\otimes\ket{n_1,n_2}$ indicates a state associated to a lattice site and $\ket{1; n_1,n_2}$,
the state associated to the corresponding basis site. The $N$-dimensional lattice subspace, ${\cal H}_P$, is spanned by kets
$\{\ket{s; n_1,n_1}\}$ with $s=0,1$.

At a given site there are three possible directions of motion. We label each of them with an
integer index $j=0,1,2$ so that the direction of motion is encoded in a three-dimensional ``coin'' subspace, ${\cal H}_C$, spanned by
$\{\ket{0},\ket{1},\ket{2}\}$. The full $3N$-dimensional Hilbert space is ${\cal H}={\cal H}_C\otimes{\cal H}_P$ and the basis
states $\ket{j; s; n_1,n_2}$ form an orthonormal set. In this basis, a generic state $\ket{\Psi}\in{\cal H}$ is expressed as
\begin{equation}\label{P-state}
  \ket{\Psi}=\sum_{j;\, n_1,n_2} a_{j,n_1,n_2} \ket{j; 0; n_1,n_2}+b_{j,n_1,n_2} \ket{j; 1; n_1,n_2}
\end{equation}
where the complex coefficients $a_{j, n_1,n_2}$ $\left(b_{j, n_1,n_2}\right)$ are the lattice (basis) components and the
normalization condition $\scalar{\Psi}{\Psi}=1$ is assumed. A step in any direction from a lattice (basis) point leads to a
basis (lattice) point, according to the propagation rule
\begin{equation}
  \ket{j; s; n_1, n_2}\rightarrow  \ket{j; \, s \oplus 1; \, n_1-(-1)^s\alpha_j,\, n_2-(-1)^s\beta_j}
\end{equation}
where $\oplus$ is the binary sum and $\mb{\hat v}_j=(\alpha_j,\beta_j)$ are the directional vectors
\begin{equation}\label{eq:dir}
  \mb{\hat v}_0 = (0,0), \quad \mb{\hat v}_1 = (1,0),\quad\mbox{and}\quad \mb{\hat v}_2 = (0,1).
\end{equation}
This conditional displacement is implemented with a shift operator,
\begin{equation}\label{Sr-op}
  S = \sum_{j,\,s,\,\mb{\hat n}} \ket{j,s\oplus 1,\mb{\hat n}- (-1)^s\,\mb{\hat v}_j}\bra{j,s, \mb{\hat n}}
\end{equation}
where we have introduced the shorthand notation $\bm{\hat n}$ for $(n_1,n_2)$ and the sum modulo $m$ is understood for
these components. The evolution operator of a quantum walk on the hexagonal network is then
\begin{equation}\label{eq:evol}
  U = S\cdot (G_3\otimes I_P)
\end{equation}
where $I_P$ is the identity in ${\cal H}_P$. The three-dimensional Grover operation $G_3$ acts in ${\cal H}_C$ and, in the
representation stated above, is given by
\begin{equation}\label{Grover}
  G_3=\frac13\left(
    \begin{array}{rrr}
      -1 & 2 & 2 \\
      2 & -1 & 2\\
      2 & 2 & -1  \\
    \end{array}\right).
\end{equation}
After $t$ iterations, an initial state $\ket{\Psi_0}$ evolves to $\ket{\Psi_t}=U^t\ket{\Psi_0}$. Note that $U$ is a real
operator, as required by the abstract search formalism~\cite{AKR05}.

For single-step displacements, the spatial part of the evolution operator is diagonal in the Fourier representation, so let
us now consider the Fourier transform in ${\cal H}_P$. The reciprocal lattice ~\cite{Kittel} to the one defined by the vectors
$\{\mb{a_1},\mb{a_2}\}$ is formed by vectors $\{\mb{g_1},\mb{g_2}\}$, which satisfy
\begin{eqnarray}\label{eq:reciprocal}
  \mb{g}_1\cdot\mb{a}_1&=&\mb{g}_2\cdot\mb{a}_2=2\pi /m,\nonumber\\
  \mb{g}_1\cdot\mb{a}_2&=&\mb{g}_2\cdot\mb{a}_1=0.
\end{eqnarray}
A point of the reciprocal lattice is located through a vector $\mb{k}=k_1 \mb{g_1}+k_2 \mb{g_2}$ for integers
$k_1,k_2\in\left[0, m-1\right]$. We shall use the short-hand notation $\mb{\hat k}$ for the two-component vector $(k_1,k_2)$.

The coin components of $\ket{\Psi}$ play no essential role in what follows, so let us for the moment omit the coin dependence. Then, a state $\ket{\Psi}$ can be expressed either in the position representation or in the wavenumber representation as
\begin{equation}\label{state}
  \ket{\Psi}=\sum_\mb{\hat n} \left( a_{\hat n} \ket{0;\mb{\hat n}}+b_{\hat n} \ket{1;\mb{\hat n}}\right)=
\sum_\mb{\hat k} \left( f_{\hat k} \ket{0;\mb{\hat k}}+ g_{\hat k} \ket{1;\mb{\hat k}}\right).
\end{equation}
The $N$ states $\ket{s;\mb{\hat k}}$ are related to the position representation by the Fourier transform
\begin{eqnarray}
  \ket{s;\mb{\hat k}}&=&\sqrt{\frac{2}{N}}\sum_\mb{\hat n} \,e^{-i\mb{k}\cdot\mb{r}} \ket{s,\mb{\hat n}},\label{F-transf}\\
  \ket{s;\mb{\hat n}}&=&\sqrt{\frac{2}{N}}\,\sum_\mb{\hat k} \,e^{i\mb{k}\cdot\mb{r}} \ket{s,\mb{\hat k}}.\label{F-inv-transf}
\end{eqnarray}
These states satisfy $\scalar{s,\mb{\hat k}}{s^\prime,\mb{\hat n}}=\sqrt{\frac{2}{N}}\,e^{i\mb{k}\cdot\mb{r}}\,\delta_{s,s^\prime}$,
so Fourier transformed kets of lattice (basis) states are orthogonal to basis (lattice) kets.

Taking into account the coin dependence and using the above relations, the action of the shift operator, eq.~(\ref{Sr-op}), on
\textbf{k}-space is
\begin{equation}
  S\ket{j;s;\mb{\hat k}}=\omega^{-(-1)^s\mb{\hat k}\cdot\mb{\hat v_j}}\,\ket{j; s\oplus 1 ;\mb{\hat k}},
\end{equation}
where $\omega\equiv\exp(2\pi i/m)$ and the directional vectors $\mb{\hat v}_j$ have been defined in eq.~(\ref{eq:dir}).  Notice that
$S$ is diagonal in $k$-space and connects lattice points with basis points as expected. This fact effectively reduces the problem to a six-dimensional subspace ${\cal L}_{\mb{k}}$ spanned by the kets $\{\ket{j;s}\}$. Since  $\mb{\hat k}$
takes $N/2$ values, the Hilbert space is now decomposed in this subspace and the one spanned by the $\ket{\mb{\hat k}}$ states,
with a dimensional count $\protect{6\times N/2=3N}$. In this six-dimensional subspace, in the representation stated above, the
reduced evolution operator $U_{\mb{k}}$ has the explicit form
\begin{equation}\label{eq:Uk-matrix}
  U_{\mb{k}}=\left(\begin {array}{cccccc} 0&-\frac{1}{3}&0&\frac{2}{3}&0&\frac{2}{3}\\\noalign{\medskip}-
      \frac{1}{3}&0&\frac{2}{3}&0&\frac{2}{3}&0\\\noalign{\medskip}0&\frac{2}{3}\,{\omega}^{{k_1}}&0&-\frac{1}{3}
      \,{\omega}^{{k_1}}&0&\frac{2}{3}\,{\omega}^{{k_1}}\\\noalign{\medskip}\frac{2}{3}\,
        {\omega}^{-{ k_1}}&0&-\frac{1}{3}\,{\omega}^{-{k_1}}&0&\frac{2}{3}\,
        {\omega}^{-{k_1}}&0\\\noalign{\medskip}0&\frac{2}{3}\,{\omega}^{{k_2}}&0&\frac{2}{3}\,
        {\omega}^{{k_2}}&0&-\frac{1}{3}\,{\omega}^{{k_2}}\\\noalign{\medskip}
        \frac{2}{3} \,{\omega}^{-{k_2}}&0&\frac{2}{3}\,{\omega}^{-{k_2}}&0&-\frac{1}{3}\,
        {\omega}^{-{k_2}}&0\end {array}\right).
\end{equation}
Its characteristic polynomial factors as
\begin{equation}
  P(\lambda)=(\lambda -1)(\lambda+1)(\lambda^4 - 2 \cos(2\theta_k) \, \lambda^2+1),
\end{equation}
where the angle $\theta_k\in [0,\frac{\pi}{2}]$ is defined by
\begin{equation}\label{eq:c2t}
  \cos(2\theta_k)\equiv  \frac{4}{9} \left(\cos \tilde{k}_1  + \cos \tilde{k}_2 + \cos\big(\tilde{k}_1-\tilde{k}_2\big)\right) - \frac{1}{3},
\end{equation}
and $\tilde{k}_i\equiv 2\pi k_i/m$ for $i=1,2$. The six eigenvalues of $U_\mb{k}$ are $\pm 1$ and $\pm e^{\pm i
  \theta_k}$.

\section{Time complexity of the search algorithm}
\label{sec:order}

The abstract search formalism described in~\cite{AKR05} provides a way to implement a spatial search algorithm on a network
where a QW has been properly defined. A convenient summary of the abstract search formalism can be found in Ref.~\cite{Tulsi08}.

Assume that the search is for a single site, $\mb{r}=\mb{r_0}$, in a periodic hexagonal (honeycomb) network with $N$ sites. The  effective target state in ${\cal H}$ is $\ket{t}\equiv\ket{u}\otimes\ket{\mb{r_0}}$ where $\protect{\ket{u}=\frac{1}{\sqrt{6}}\sum_{j,s} \ket{j,s}}$ is the uniform superposition in ${\cal L}_\mb{k}$.

The generalized search algorithm iterates the unitary operator
\begin{equation}\label{eq:gen-search}
  U^\prime=U\cdot R_t,
\end{equation}
where  $U$ is the unperturbed quantum walk operator defined in eq.~(\ref{eq:evol}) and $R_t\equiv I_{3N}-2\op{t}$. In the introduction, we mentioned that a generalized search is implemented with a modified quantum walk operator of the form $U'=S\cdot C'$, where $C'$ is a unitary coin operation that acts differently on the searched site, \textit{i.e.} $C'=C\otimes (I_P -\op{\mb{r_0}})+C_1\otimes \op{\mb{r_0}}$. Both forms for $U^\prime$ are equivalent, provided the Grover coin $C=G_3$ is used and the usual choice of $C_1=-I_C$ is made for the coin operation on a searched site.

The initial state for the algorithm is the uniform superposition in ${\cal H}$,
\begin{equation}
  \ket{\Phi_0}  = \ket{u}\otimes\ket{u_P}=\frac{1}{\sqrt{3N}}\sum_{j,s,\mb{\hat n}} \ket{j;s;\mb{\hat n}},
\end{equation}
where $\ket{u_P}\equiv\sqrt{\frac{2}{N}}\sum_{\mb{\hat n}}\ket{\mb{\hat n}}$ is the uniform superposition in position space. Except for a phase shift, the operator $R_t$ implements a reflection about the effective target $\ket{t}$ and a single application of $R_t$ on the uniform superposition ``marks'' the searched state by changing its relative phase, in a similar form as in Grover's search algorithm \cite{Gro96a}.

As mentioned previously, Ambainis et al. prove the remarkable result that the time complexity of the abstract search algorithm depends on the eigenproblem of $U$ alone~\cite{AKR05}. They show that, after $T=O(1/\alpha)$ iterations of $U_A$, the initial state evolves to a final state $\ket{\Phi_f}=U_A^T\ket{\Phi_0}$ which has an increased overlap $|\scalar{\Phi_f}{t}|$ with the effective searched state $\ket{t}$. Detailed expressions for the dependence of $\alpha$ and $\scalar{\Phi_f}{t}$ on the eigenvalues and eigenvectors of $U$ are given below. The unperturbed operator $U$ must satisfy two conditions: (i) $U$ must be a real operator and (ii) the uniform superposition state $\ket{\Phi_0}$ must be a non-degenerate eigenstate of $U$ with eigenvalue $1$. Both conditions are met by the quantum walk operator $U$ defined in eq.~(\ref{eq:evol}), since $G_3$ is real and $(G_3\otimes I_2)\ket{u}=\ket{u}$.


We follow the notation of Ref.~\cite{Tulsi08} to describe the eigenproblem for $U$. The eigenvectors associated with the $-1$ eigenvalue, which may be $M$-degenerate, are labeled as $\ket{\Phi_i}$ for $i=1\ldots M$. Let $\ket{\Phi_\ell^\pm}$ indicate the eigenvectors associated to all other eigenvalues distinct from $\pm1$. The eigenvectors may be chosen so that the amplitudes on $\ket{t}$ on the
proper basis of $U$ are real. Then, the effective target state may be expanded with real coefficients as
\begin{equation}\label{eq:t-expan}
  \ket{t}=a_0\ket{\Phi_0} + \sum_\ell a_\ell(\ket{\Phi_\ell^+ + \Phi_\ell^-}) + \sum_{i=1}^M a_i\ket{\Phi_i},
\end{equation}
where the index $\ell$ runs over all pairs of conjugate eigenvectors with eigenvalues distinct from $\pm 1$.
These amplitudes $(a_0,a_\ell,a_i)$, together with the angles $\theta_k$ defined by eq.~(\ref{eq:c2t}), determine the time complexity
of the abstract search algorithm \cite{AKR05,Tulsi08}. The rotation angle towards the searched element, which results
from a single application of $U^\prime$, is
\begin{equation}\label{eq:alpha}
  \alpha=O\left(a_0\left[\sum_\ell \frac{a_\ell^2}{1-\cos\theta_\ell}+\frac14\sum_{i=1}^M a_i^2\right]^{-\frac12}\right).
\end{equation}
After $T=\pi/2\alpha$ iterations, the overlap with the searched state
is
\begin{equation}\label{eq:overlap}
  \left|\scalar{t}{\alpha^+}\right|=O\left(\min \left[\left(\sum_\ell a_\ell^2\cot^2\frac{\theta_\ell}{4}\right)^{-\frac12},1\right]\right).
\end{equation}
In both expressions, the sums $\sum_\ell$ run over the eigenvalues distinct from $\pm 1$.

The (unnormalized) eigenvectors  $\ket{\nu^{\pm}_\mb{k}}$ associated with the eigenvalues $\pm 1$ are
\begin{equation}
  \ket{\nu^{\pm}_\mb{k}}\propto\left( \begin {array}{c} \pm({\omega}^{k_{{1}}}-{\omega}^{k_{{2}}})\\
      \noalign{\medskip}{\omega}^{k_{{2}}}-{\omega}^{k_{{1}}}\\
      \noalign{\medskip} \pm{\omega}^{k_{{1}}} \left({\omega}^{k_{{2}}}-1\right)\\
      \noalign{\medskip}1-{\omega}^{k_{{2}}}\\
      \noalign{\medskip}\pm{\omega}^{k_{{2}}} \left(1-{\omega}^{k_{{1}}} \right)\\
      \noalign{\medskip}{\omega}^{k_{{1}}}-1\end {array} \right)
\end{equation}
except for $k_1=k_2= 0$. Note that the projection of the effective target state $\ket{t}$ on ${\cal L}_\mb{k}$ is the uniform state
$\ket{u}$ and $\scalar{u}{\nu^\pm_\mb{k}}=0$, unless $k_1=k_2=0$.
In this degenerate case, the eigenvalues are $\pm 1$ and $\ket{u}$ itself is an eigenvector of $U_\mb{k}$ with eigenvalue
$+1$. All the other eigenvectors are orthogonal to $\ket{u}$ so, for
all $\mb{k}$,
\begin{equation}
  \scalar{u}{\nu^+_\mb{k}}=\delta_{\mb{k},\mb{0}}\quad\mbox{and}\quad\scalar{u}{\nu^-_\mb{k}}=0
\end{equation}
so that $a_0=\sqrt{2/N}$ and the terms corresponding to the eigenvalue $-1$ do not contribute in eq.~(\ref{eq:t-expan}).
Let us indicate the eigenvectors associated to the other eigenvalues $\pm e^{\pm i\theta_k}$ as $\ket{\pm\nu_\mb{k}^{(\pm\theta_k)}}$.
Then, eq.~(\ref{eq:t-expan}) for the effective target state reduces to
\begin{multline}
  \ket{t}=\sqrt{\frac{2}{N}}\ket{u,u_P}+\sqrt{\frac{2}{N}}\sum_{\mb{k}\ne\mb{0}}\left[ a_\mb{k}^+
           \left(\ket{+\nu_\mb{k}^{(\theta_k)}}+\ket{+\nu_\mb{k}^{(-\theta_k)}} \right)+ \right.\\  \left.
    a_\mb{k}^-\left(\ket{-\nu_\mb{k}^{(\theta_k)}}+\ket{-\nu_\mb{k}^{(-\theta_k)}} \right)\right]\otimes\ket{\mb{k}},
\end{multline}
with the real amplitudes
\begin{equation}\label{eq:ak1}
  a_\mb{k}^{\pm} = \frac12 \,\sqrt{1\pm \frac{1+\cos\tilde k_1 + \cos\tilde k_2}{3\cos\theta_k}}.
\end{equation}
Even though analytical expressions for all the eigenvectors of $U$ are unknown, knowledge of the coefficients
$a_\mb{k}^{\pm}$ allows us to evaluate the time complexity of the search algorithm.

For the quantum walk on a honeycomb, eq.~(\ref{eq:alpha}) leads to
\begin{equation}\label{eq:AN}
  \frac{1}{\alpha}=O\left(\sqrt{\sum_{\mb{k}\neq \mb{0}}\frac{(a_\mb{k}^+)^2}{1-\cos\theta_k}+\frac{(a_\mb{k}^-)^2}{1+\cos\theta_k}}~\right)\equiv O\left(\sqrt{A(N)}\right).
\end{equation}
Let us concentrate on the $N$-dependence, for $N\gg 1$, of the argument $A(N)$ of the above square root. Using eq.~(\ref{eq:ak1}), after some
manipulation, we obtain
\begin{equation}\label{eq:integra1}
  A=\frac{1}{6}\sum_{\mb{k}\neq 0}\frac{4+\cos\tilde k_1+\cos\tilde k_2}{\sin^2\theta_k}\approx\frac{N}{48}
\frac{1}{(\pi-\varepsilon)^2} \iint_\varepsilon^{2\pi-\varepsilon}d\tilde k_2 d\tilde k_1 \frac{4+\cos\tilde k_1+\cos\tilde k_2}{\sin^2\theta_k}.
\end{equation}
where we have used $\protect{\sin^2\theta_k = \frac23
  -\frac29\left(\cos\tilde k_1+\cos\tilde k_2+\cos(\tilde k_1-\tilde k_2)\right)}$ and approximated the sum by an integral in the usual form, $\sum_{\mb{k}\neq 0}\rightarrow\frac{N}{8}\frac{1}{(\pi-\varepsilon)^2}\iint_\varepsilon^{2\pi-\varepsilon}
d\tilde k_1 d\tilde k_2$ with $\protect{\varepsilon=2\pi\sqrt{2/N}}$. For $N\gg 1$ (or $\protect{\varepsilon\ll 1}$), the $N$-dependence of $A$ is
$$
A(N)\simeq
\frac{3N}{32}\frac{1}{\pi^2}\int_\varepsilon^{2\pi-\varepsilon}
d\tilde k_2 \int_\varepsilon^{2\pi-\varepsilon} \frac{d\tilde k_1}{\tilde k_1^2 + \tilde k_2^2 - \tilde k_1\tilde k_2}
\sim N\log\left(\frac{2\pi}{\varepsilon}\right)\sim N\log N.
$$
So, $1/\alpha=O(\sqrt{N\log N})$ iterations of $U^\prime$ are required to reach the final state $\ket{\Phi_f}$.

Using eq.~(\ref{eq:overlap}), we obtain that the inverse of the overlap between
the final state and the target $\ket{t}$ is
\begin{equation}
  {\left|\scalar{t}{\Phi_f}\right|}^{-2}=
O\left({\frac{2}{N} \sum_{\mb{k}\neq 0} \left[\left(a_k^+\right)^2+\left( a_k^-\right)^2\right]\cot^2\left(\theta_k/4\right)}\right)\equiv O\left({B(N)}\right).
\end{equation}
Using eq.~(\ref{eq:ak1}) and for $N\gg 1$, the $N$-dependence of $B(N)$ is
\begin{equation}
  B(N)\simeq\frac{1}{N}\sum_{\mb{k}\neq 0}\cot^2\left(\theta_k/4\right)\simeq\frac{1}{8(\pi-\varepsilon)^2}
\iint_\varepsilon^{2\pi-\varepsilon}d\tilde k_2 d\tilde k_1 \cot^2\left(\frac{\theta_k}{4}\right)\sim \log N,
\end{equation}
where the divergence comes, as before, from the $\sin^{-2}\theta_k$ term. Then
\begin{equation}
  \frac{1}{\left|\scalar{t}{\Phi_f}\right|^2}=O\left({\log N}\right).
\end{equation}

The analysis of the time complexity of the algorithm is as follows. After $1/\alpha=O(\sqrt{N\log N})$ iterations of $U^\prime$, the algorithm reaches the final state $\ket{\Phi_f}$ with probability $p={\left|\scalar{t}{\Phi_f}\right|^2}$. The method known as amplitude amplification~\cite{BHMT02}
states that if there is an unitary operator $U'$ such that the probability of measuring a marked 
state upon measuring $U'^t\ket{\Phi_0}$ is $p>0$, then there is a quantum procedure that finds the marked 
state with certainty using $O(1/\sqrt p)$ applications of $U'^t$. That procedure uses the inversion about the 
mean, which can be implemented in $O(\sqrt N)$ steps. This leads to an overall complexity of $O(\sqrt{N}\,\log N)$ to find the marked state in the honeycomb lattice. This is the same complexity of the AKR spatial-search algorithm on the cartesian grid of a torus~\cite{AKR05}, where each site has four neighboring sites and the $N$ sites form a lattice.

\begin{figure}[h]
\centering
  \setlength{\unitlength}{0.65pt}
  \begin{picture}(160,160)(80,0)
    \put(43,25){\makebox(0,0)[r]{$\vert{u_{s}}\rangle$}}
    \put(245,125){\line(1,0){25}}
    \put(80,115){\framebox(20,20){$X_{\delta}$}}
    \put(150,115){\framebox(20,20){$X_{\delta}^{\dagger}$}}
    \put(220,115){\framebox(25,20){$-Z$}} \put(218,25){\line(1,0){52}}
    \put(218,75){\line(1,0){52}}
    \put(102,15){\framebox(46,70){\small{$R_t$}}}
    \put(148,25){\line(1,0){24}} \put(148,75){\line(1,0){24}}
    \put(195,125){\circle*{8}} \put(195,121){\line(0,-1){36}}
    \put(172,15){\framebox(46,70){$U$}}
    \put(43,125){\makebox(0,0)[r]{$\vert{1}\rangle$}}
    \put(43,75){\makebox(0,0)[r]{$\vert{u_{c}}\rangle$}}
    \put(125,125){\circle*{8}} \put(170,125){\line(1,0){50}}
    \put(100,125){\line(1,0){50}} \put(55,125){\line(1,0){25}}
    \put(125,121){\line(0,-1){36}} \put(55,25){\line(1,0){47}}
    \put(55,75){\line(1,0){47}}
  \end{picture}
  \caption{Tulsi's circuit diagram for the one-step evolution operator of the quantum walk search algorithm.}\label{fig:circ}
\end{figure}
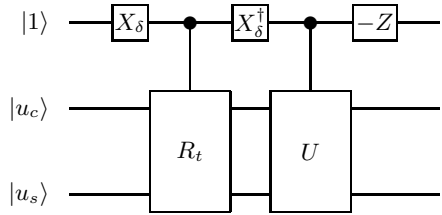

In a remarkable paper, A.~Tulsi~\cite{Tulsi08} has described a method to improve even further the probability of finding the marked vertex.
Let us consider a quantum circuit that implements operator $R_t$ followed by $U$ as defined in eq.~(\ref{eq:gen-search}).
Tulsi introduced an extra qubit and defined a new one-step evolution operator as described in the circuit of
Fig.~\ref{fig:circ}, where $-Z$ is the negative of Pauli $Z$ operator
and
\begin{equation}
  X_{\delta}= \left( \begin{array}{cc} \cos\delta & \sin\delta \\ -\sin\delta & \cos\delta\end{array} \right),
\end{equation}
where $\delta$ must assume the value $1/\sqrt{\log N}$.
It is straightforward to show that Tulsi's procedure increases the overlap between the final state and the target, such that $\protect{|\scalar{t}{\Phi_f}|=O(1)}$.
Consequently, the overall time complexity of the search algorithm in the honeycomb lattice may be improved to $O(\sqrt{N \log N})$, as in the AKR case, with Tulsi's modification. It is not necessary to use the amplitude amplification method in this case.  We have performed an independent numerical simulation which agrees with this analytical calculation.

\section{Conclusions}
\label{sec:conc}

Hexagonal networks (honeycombs) are the underlying representation of a carbon structure called graphene, which has been attracting special attention over the last years, especially for its potential applications in nanotechnology. In this paper, a new quantum algorithm for spatial search in a honeycomb with periodic boundary conditions is discussed. The protocol is based on a quantum walk in the honeycomb. We obtain the expression for the evolution operator in the Fourier representation and solve its eigenvalue problem. Then, the abstract search formalism developed by Ambainis et al. \cite{AKR05} is used to obtain the complexity of the algorithm from the partially known spectral decomposition of the evolution operator. Our results have been verified by numerical simulations.

The search algorithm on the honeycomb has an overall time complexity of $O(\sqrt{N}\log N)$ by using the amplitude amplification procedure. A better improvement, to  $O(\sqrt{N\log N})$, can be obtained by using Tulsi's technique. Surprisingly, this is the same complexity found for the quantum search on the square grid after Tulsi's improvement. Both the hexagonal grid and the square grid are regular graphs which cover the plane, although the former has degree~$3$ and the latter has degree~$4$.  The fact that the complexity of the search algorithm is the same in both cases suggests that the number of connections of each node is not affecting the complexity of the abstract spatial search algorithm.

Several open questions remain. One of them is whether the abstract search algorithm has the same complexity when applied to graphs of general degrees. The triangular network, for instance, has degree~$6$ and also covers the plane. It would be interesting to investigate the behavior of the algorithm on this topology. One may also inquire about how robust the search algorithm is when there are some missing nodes. Finally, we point out that an optimal spatial search algorithm $O(\sqrt{N})$ for the case of a two-dimensional network covering the plane has not yet been found.

\section*{Aknowledgements} We acknowledge helpful discussions with R.~Marotti and R.~Siri and thank M. Forets for help in revising the final version of the manuscript.  This work was done with financial support from PEDECIBA (Uruguay) and CNPq (Brazil).

\end{document}